\documentstyle[prl,epsf,amsbsy,amstext,amsfonts,amssymb,aps]{revtex}

\begin{document}

\title{Electroproduction of strangeness above the resonance region.}
\author{M. Guidal $^{a}$, J.-M. Laget $^{b}$ and M. Vanderhaeghen$^{c}$}
\address{$^{a}$ IPN Orsay, F-91406 Orsay, France}
\address{$^{b}$ CEA/Saclay, DAPNIA/SPhN, F-91191 Gif-sur-Yvette Cedex,
France}
\address{$^{c}$ University Mainz, D-55099 Mainz, Germany}
\date{\today}
\maketitle

\begin{abstract}
A simple and elegant model, based on Reggeized $t$-channel exchanges
is successful in reproducing the main features of all existing data
of the reactions $ep\to e^\prime K^+\Lambda$ and
$ep\to e^\prime K^+\Sigma$. In particular, the original way
gauge invariance is taken into account is found to be essential to describe
the ratio between the Coulomb and the transverse cross-sections at large
$Q^2$ that has been measured recently at JLab.
\end{abstract}


\twocolumn

Strangeness production is undergoing a  renewed interest in view of  the
numerous data which are  currently coming out of  electron accelerators
like CEBAF  at Jefferson  Lab and  ELSA at  Bonn.   It offers us with an
original  way  to  probe  hadronic  matter.    Not only the study of the
Hyperon-Nucleon interaction is  a mandatory complement  to the study  of
the Nucleon-Nucleon  interaction, but  the implantation  of an  impurity
(the strange quark or a hyperon)  in a hadronic system is a  formidable
tool to  study its  properties.   However, the  elementary processes  of
photo- and electroproduction  of kaons off the nucleon should be 
mastered. In particular, we will show that the determination of the
kaon form factor depends on the model used.\par
At low energy, within about 1~GeV above threshold, many resonances  may
contribute in the $s$-channel and  fits to the scarce data  are generally
obtained       at       the       expense       of       many       free
parameters~\cite{bobw},\cite{david}.      At   higher   energy,    Regge
phenomenology provides us with an elegant and simple way to account  for
the analyticity and  unitarity of the amplitude with  almost no free
parameters~\cite{paper},\cite{article}.

\begin{figure}[h]
\epsfxsize=8.5 cm
\epsfysize=10. cm
\centerline{\epsffile{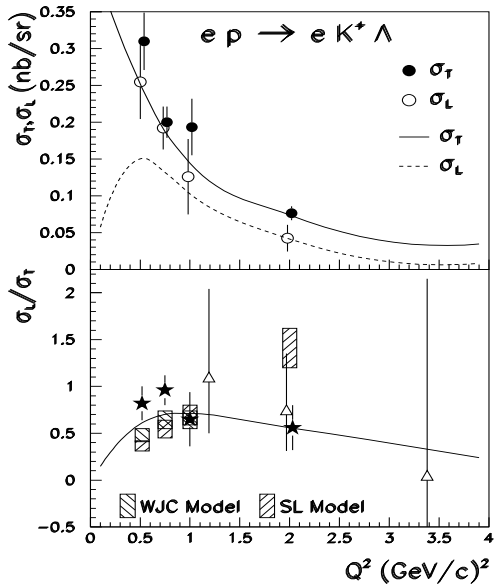}}
\caption[]{$Q^2$ dependence of $\sigma_T$ and $\sigma_L$ (upper plot) and
$\sigma_L /\sigma_T$ (lower plot) for the $\gamma^* + p \rightarrow K^+ +
\Lambda$ reaction, at W = 1.84 GeV. Experimental data points are from~:
($\bullet$, $\circ$, $\bigstar$)~: \cite{baker} and ($\triangle$)~: 
\cite{bebek77b}. Besides our calculation, on the lower panel are also 
shown the models of Ref.~\cite{bobw},\cite{david}. The WJC prediction is $\approx$ 3.5 (off scale) for $Q^2$ =2.0 (GeV/c)$^2$.}
\label{fig:baker}
\end{figure}

Our Regge model is fully described
in Refs.\cite{paper},\cite{article}  for
the {\it photoproduction} of pions and kaons on the nucleon above
the resonance region (above $E_{\gamma} \approx$ 2 GeV), and in
Ref.\cite{electro} for the {\it electroproduction} of pions. 
It is based simply on two ``Reggeized" $t$-channel exchanges ($\pi$ and
$\rho$
trajectories for pion production, $K$ and $K^*$ trajectories for kaon
production). An original and essential feature of this model is the way gauge
invariance is restored for the $\pi$ ($K$) $t$-channel exchanges by proper
``reggeization" of the $s$-channel nucleon pole contribution. This was found to
be the key element to describe numerous features of the experimental data
(for instance, the $\pi^+$/$\pi^-$ ratio, the forward peaking of the charged
pion differential cross section, the photon asymmetry, etc...).\par

As in Ref.\cite{electro}, we extend the model of kaon photoproduction
to electroproduction by multiplying the separately gauge invariant $K$ and
$K^*$ $t$-channel diagrams by a monopole form factor~:
\begin{equation}
F_{K, K^*}(Q^2) \;=\; \left[ 1 + Q^2/\Lambda_{K, K^*}^2 \right]^{-1} \;,
\label{eq:kff}
\end{equation}
with $Q^2 = -q^2$, where $q$ is the spacelike virtual photon four-momentum.
The mass scales $\Lambda_{K}$ and $\Lambda_{K^*}$ are chosen to be
$\Lambda^2_K= \Lambda^2_{K^*} = 1.5$ GeV$^2$, in order to fit the high
$Q^2$ behavior of $\sigma_L$ and $\sigma_T$ in Fig.~\ref{fig:baker}. We
keep the same coupling constants at the
($K,(\Lambda,\Sigma),N$) and
($K^*,(\Lambda,\Sigma),N$) vertices as in the photoproduction study
\cite{article},
which were found to describe all existing high energy data, i.e.~:


\begin{eqnarray}
&&\frac{g^2_{K\Lambda N}}{4\pi}=10.6 \; , \; g_{K^*\Lambda N}=-23.0 \; , 
\; \kappa_{K^*\Lambda N}=2.5, \nonumber\\
&&\frac{g^2_{K\Sigma N}}{4\pi}=1.6 \; \; \; , \;  g_{K^*\Sigma N}=-25.0 \; ,
\; \kappa_{K^*\Sigma N}=-1.0
\label{ctes}
\end{eqnarray}
The magnitudes and signs ($g_{K\Lambda N}<0$ and $g_{K\Sigma N}>0$) of 
the $K$ strong coupling constants are in agreement with SU(3) constraints 
(broken to about 20$\%$). The signs of the $K^*$ strong coupling constants 
are also in accordance with SU(3).\par

It turns out that the recent measurement~\cite{baker}, at Jefferson Lab,
of  the  ratio  between  the  Coulomb  ($\sigma_L$)  and  the transverse
($\sigma_T$) cross-sections of the $p(e,e'K^+)\Lambda$ reaction  clearly
favors the Regge model over  the resonance models, already in  the CEBAF
energy range.   Fig.~\ref{fig:baker} shows the  comparison of the  Regge
model with the data.  Particularly interesting is the behavior at  large
$Q^2$ of the $\sigma_L/\sigma_T$ ratio  which decreases as the data,  in
contrast      to       two      recent       theoretical      resonance
models~\cite{bobw},\cite{david}. Our Regge model suitably reproduces the
trend of this ratio.  The  reason is that, due to gauge  invariance, the
$t$-channel kaon  exchange and  the $s$-channel  nucleon pole  terms are
indissociable  and  must  be  treated  on  the  same  footing.    In  our
model~\cite{paper},\cite{article}, they  are Reggeized  in the  same way
and multiplied by the same  electromagnetic form factor.  This  approach
clearly differs from  traditional ones \cite{bobw},\cite{david}  where a
different electromagnetic form factor is assigned  to the $t$- and 
$s$-channel  diagrams
(monopole  and  a  dipole  forms  respectively -with also different mass
scales-, in general).  This  explicitly breaks gauge invariance and  the
introduction of  a purely  phenomenological and  ad-hoc counter-term  is
needed to restore it.   Whatever particular way this procedure has been done
in the existing literature, it produces a -linearly- rising ratio 
$\sigma_L /\sigma_T$ with $Q^2$ in contrast to the data. It is important to 
note that the $Q^2$ dependence of this ratio is relatively insensitive to the 
particular shape or mass scale taken for the electromagnetic form factors 
of the $K$ and $K^*$ and that the decreasing trend observed here is 
intrinsically linked to the assignment of the {\it same} electromagnetic 
form factor to the $t$- and $s$-channel Born diagrams (which is clearly 
the simplest way in fact to keep gauge invariance in electroproduction).\par

\begin{figure}
\epsfxsize=8.5 cm
\epsfysize=11. cm
\centerline{\epsffile{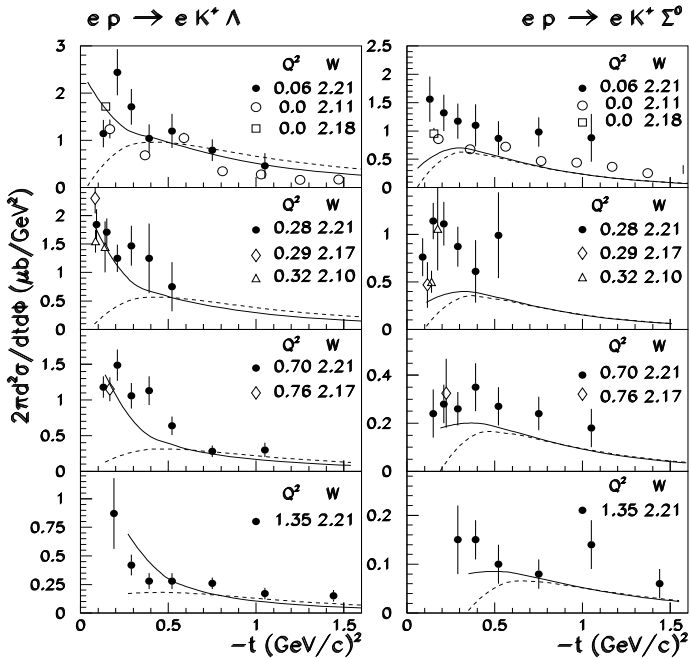}}
\caption[]{$t$ dependence of the $\gamma^* + p \rightarrow K^+ + \Lambda$
(left) and $\gamma^* + p \rightarrow K^+ + \Sigma^0$ (right) differential
electroproduction cross section $2\pi d^2\sigma/dtd\Phi$ for different
$Q^2$ values. The solid line shows the $K$+$K^*$ exchanges, the dashed
line shows the $K^*$ exchange only. Displayed data correspond approximately
to the same $W$ and $Q^2$ ranges (units for the figure are in GeV/c and
(GeV/c)$^2$ respectively )~; ($\bullet$)~: \cite{brauel79}, 
($\circ$)~: \cite{bonn}, ($\Box$)~:
\cite{feller}, ($\diamondsuit$)~: \cite{bebek77}, ($\triangle$)~:
\cite{azemoon}.}
\label{fig:brauel}
\end{figure}

The Regge model reproduces also fairly well the -scarce- data prior to 
Jefferson Lab. Fig.~\ref{fig:brauel}
shows the $t$-dependence of the $\gamma^* + p \rightarrow K^+ + \Lambda$
and $\gamma^* + p \rightarrow K^+ + \Sigma^0$ differential
electroproduction cross section $2\pi d^2\sigma/dtd\Phi$ for different
$Q^2$ values. The latest~\cite{bonn} and older~\cite{feller} Bonn data for
photoproduction are also shown for reference.
At $Q^2$=0.06 GeV$^2$, there is essentially no influence
of the form factors. Therefore, without any additional parameter,
a straightforward extension of the photoproduction
model gives the correct $t$-dependence and magnitude of the data.
As in the photoproduction study, the
$\Lambda$ and $\Sigma$ channels show a different behavior at forward
angles~: the differential cross section decreases towards 0
for the latter one whereas it tends to peak for the former one.
According to~\cite{article},  this ``peaking" for the $\Lambda$ channel
is due to the dominance of the gauge invariant $K$ exchange at small $t$. Because of
the
weaker $g_{K\Sigma N}$ coupling constant relative to the $g_{K\Lambda N}$
coupling constant, the $K^*$ exchange contribution -which has to vanish at forward angles
due to angular momentum conservation- dominates the $\Sigma$ channel,
which reflects into a decrease of the differential cross section
at forward angles. This decrease at small $t$ is attenuated at larger $Q^2$
due to the ``shift" of $t_{min}$ with $Q^2$.\par

The Brauel et al. data of Fig.~\ref{fig:brauel} were integrated in
$\Phi$
between 120$^o$ and 240$^o$ \cite{brauel79} ; so was the model in order
to correctly take into account the influence of the
$\sigma_{TT}$ and $\sigma_{TL}$ terms which is found not to be negligible.
Fig.~\ref{fig:brauel} shows furthermore the destructive interference between 
the $K$ and $K^*$
exchange mechanisms for the $\Sigma$ channel found at large angles,
which was also noticed before in the photoproduction study \cite{article}.\par

\begin{figure}[h]
\epsfxsize=7.5 cm
\epsfysize=9. cm
\centerline{\epsffile{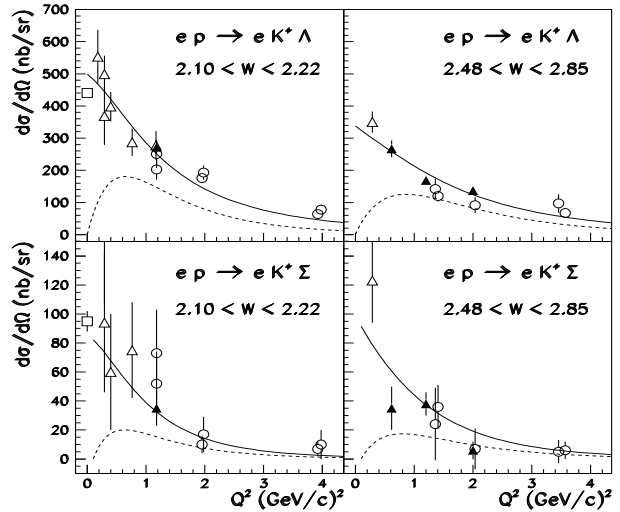}}
\caption[]{$Q^2$ dependence of the $\gamma^* + p \rightarrow K^+ + \Lambda$
(upper plots) and $\gamma^* + p \rightarrow K^+ + \Sigma^0$ (lower plots)
differential cross section $d\sigma/d\Omega$ for 2 energy bins.
Solid lines~: $d\sigma_{T+L}/d\Omega$, dashed line~: $d\sigma_{L}/d\Omega$.
Displayed data correspond approximately to the same
$W$ and $\theta$ ranges~; ($\triangle$)~: \cite{brown}, ($\circ$)~:
\cite{bebek77}, ($\blacktriangle$)~: \cite{bebek74},
($\Box$)~: \cite{feller}. This latter point has been renormalized (see text
for details).}
\label{fig:q2dep}
\end{figure}

This $Q^2$ dependence is confirmed by Fig.~\ref{fig:q2dep} which shows the
differential cross section $d\sigma/d\Omega$ at $\theta_{c.m.}$=8$^o$ as
a function of $Q^2$ for 2 energy bins. In fact, commonly,
this observable has been plotted at a single averaged $W$ value ($<W>$=2.15 GeV)
where a $\mid p^*_K\mid /W/(s-m_p^2)$ dependence was used for the
extrapolation of the lower and higher measured $W$ values \cite{bebek77}. Fig.~\ref{fig:q2dep}
shows that this procedure is approximately right for the $\Lambda$ channel
which shows roughly a $\frac{1}{s}$ behavior, but is not appropriate for the
$\Sigma$ channel which shows a rather constant behavior in this energy domain.
 Indeed, it is well known that a Regge amplitude proportional to
$s^{\alpha(t)}$ leads to a differential cross-section
$\frac{d\sigma}{dt}\propto s^{2\alpha(t)-2}$ and therefore
$\frac{d\sigma}{d\Omega}\propto s^{2\alpha(t)-1}$. For a $K$-meson exchange
dominated mechanism (such as the $\Lambda$ channel at forward angles,
see Fig.~\ref{fig:brauel}) with $\alpha_K(0)\approx$ -0.17, this implies
$\frac{d\sigma}{d\Omega}\propto s^{-1.34}$.
And for a $K^*$-meson exchange dominated mechanism (such as the
$\Sigma$ channel) with $\alpha_{K^*}(0)\approx 0.25$, we have
$\frac{d\sigma}{d\Omega}\propto s^{-0.5}$.\par

Note that the photoproduction point of \cite{feller}
has been renormalized. This data was taken at $\theta_{c.m.}$=25$^o$
($t\approx$-0.15 GeV$^2$). It has been extrapolated to 8$^o$
($t\approx$-0.06 GeV$^2$) to
allow a consistent comparison with the other data. We did so by using
the $t$-dependence of our model. From Fig.~\ref{fig:brauel}, it can be seen
that this implies an {\it upscaling} (of $\approx$ 1.2) of the $\Lambda$
photoproduction point and a {\it downscaling} (of $\approx$ 2.4) of the
$\Sigma$ photoproduction point. This leads to a very different figure
and conclusion than in Ref.~\cite{bebek77} (see Figs.~6 and 7 of Ref.~\cite{bebek77})~: firstly,
the $Q^2$ dependence of the $\Sigma$ channel is not steeper than for the
$\Lambda$ channel and, secondly, there is no particular evidence of a rise
with $Q^2$ of the $\Lambda$ cross-section which would have indicated a strong
contribution of $\sigma_L$. This latter contribution is seen to account
for less than half of the cross section.\par

It   is   remarkable   that    the   value   of   the    cut-off   mass,
$\Lambda^2_K$=$\Lambda^2_{K^*}$=1.5 GeV$^2$, deduced from the  Jefferson
Lab experiment leads  also to a  correct $Q^2$ dependence  for the world
set of previous data, both in the $\Lambda$ and the $\Sigma$ channels. It
appears, however,  to be  quite large, resulting in a rather flat form 
factor.   Indeed,  the effective  charge
radius    corresponding    to    $\Lambda^2_{K}$=1.5    GeV$^2$     is~:
$<r^2_{K}>\equiv          -6          \frac{dF_{K}}{dQ^2}\mid_{(Q^2=0)}=
\frac{6}{\Lambda^2}\approx$0.16 fm$^2$.  This  has to be compared  to the
value measured by direct scattering of kaons on atomic electrons at  the
CERN SPS which yielded $<r^2_{K}>$=0.34 fm$^2$ \cite{amen}.  However, we found previoulsy~\cite{electro}  a
good   agreement   between   the    pion   form   factor   mass    scale
($\Lambda^2_\pi=$0.462 GeV$^2$, i.e. $<r^2_{\pi}>$= 0.52 fm$^2$) deduced from  a
study similar  to the  one in  this article  and the  value measured  by
direct  scattering  of  pions  on  atomic  electrons  at  the CERN SPS~:
$\approx$0.44 fm$^2$ \cite{amen2}.  An interpretation of this discrepancy
for the kaon case can be that, the kaon pole being far from the physical
region, the form factor  used in this kind  of model does not  represent
the properties of the kaon itself but rather the properties of the whole
trajectory. Instead of being sensitive 
to the kaon form factor, one might in fact measure a transition form 
factor between the kaon and an orbital excited state lying on the kaon
Regge trajectory.
This has to be kept in mind when trying to extract the kaon
electromagnetic  form  factor  from  these electroproduction reactions.

It is clear that the extracted form factor mass scale depends
strongly on the extrapolation from $t_{min}$ to the pole. Indeed, the 
standard procedure uses a $\frac{1}{t-m^2_K}$ dependence 
reflecting the traditional $t$-channel kaon propagator. However, in our 
present approach, this propagator is proportional to $s^{\alpha(t)}$ 
which leads 
to a steeper (exponential) $t$ dependence. It is therefore no wonder 
intuitively that the mass scale of the electromagnetic form factor in this 
latter approach is softer than when using a standard Feynman propagator for
the extrapolation.\par

\begin{figure}[h]
\epsfxsize=7. cm
\epsfysize=9. cm
\centerline{\epsffile{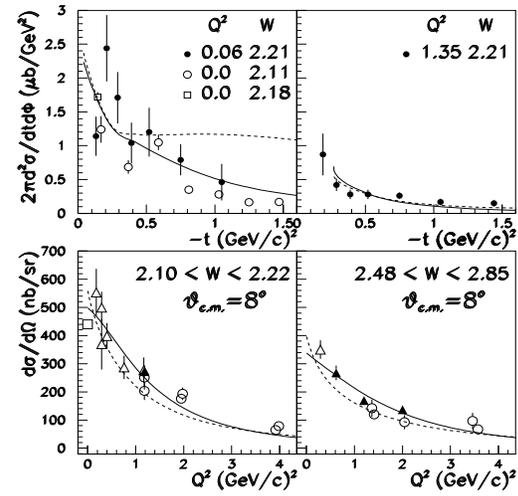}}
\caption[]{Comparison of the Regge model -solid lines- with a Born model 
-dashed lines- (see text for details) for $\gamma^* + p \rightarrow K^+ 
+ \Lambda$. Upper plots~: $t$ dependence of the differential cross sections 
at $Q^2$=0.0-0.6 GeV$^2$ (left) and $Q^2$=1.35 GeV$^2$ (right). Lower plots~: 
$Q^2$ dependence of the forward differential cross sections for 
$<$W$>$=2.16 GeV (left) and $<$W$>$=2.66 GeV (right). Experimental points 
as in Figs.~\ref{fig:brauel} and~\ref{fig:q2dep}.}
\label{fig:poleregge}
\end{figure}

To understand this crucial point better, we compare on 
Fig.~\ref{fig:poleregge} the two approaches~: the plots are extracted 
from Fig.~\ref{fig:brauel} and~\ref{fig:q2dep} and we compare here our 
Regge model (solid line) and a standard Born model based on the usual 
$\frac{1}{t-m^2_K}$ (dashed line). Our Regge model contains $K$ and $K^*$
trajectories exchanges but, as can be seen from Fig.~\ref{fig:brauel}, only
the $K$ trajectory contributes significantly. Our Born model here uses only
a (gauge-invariant) $K$ exchange rescaled in order to match the photoproduction
result in the forward direction. We left out the $K^*$ exchange as it is well known that it diverges with rising energy, due to derivative couplings 
for exchanged high spin particles. The upper left plot of 
Fig.~\ref{fig:poleregge} shows the $t$ dependence of the differential cross section at 
(almost) the photoproduction point. We see that the Born model based on $K$ exchange 
alone produces a flatter 
$t$-dependence at larger $t$ than the Regge model and the data. This could be corrected 
in principle by introducing an extra {\it hadronic} form factor at the 
$K\Lambda N$ vertex but at the expense of one additional free parameter for the corresponding mass scale. However, this will not give the correct high 
energy Regge dependence ($\frac{d\sigma}{d\Omega}\propto s^{2\alpha(t)-1}$
and the associated ``shrinkage") 
as was illustrated previously for the pion case ~\cite{electro} where 
more data are available. A decent 
$Q^2$ dependence for this Born model is obtained with an
electromagnetic monopole form factor with $\Lambda^2_K$=0.68 which is the value exactly corresponding 
to the kaon charge radius, and, in any case, much smaller than 
the value needed for the Regge model ($\Lambda^2_K$=1.5). It is clear 
that it is at small $Q^2$ values that one is 
most sensitive to the mass scale of the electromagnetic form factor
as, at larger $Q^2$, both form factors show a $\frac{1}{Q^2}$ 
asymptotic behavior. The conclusions here are clear~: a traditional
Born model (with a $\frac{1}{t-m^2_K}$ standard Feynman propagator)
seems to lead at first sight to a mass scale for the kaon electromagnetic 
form factor ($\Lambda^2_K$=0.68) compatible with the kaon charge radius. 
However, such Born model is unable to reproduce the correct energy and
$t$ dependences (unless, for this latter case, corrected by an extra {\it hadronic} 
form factor). Furthermore, it is unable to take into account
the role of the $K^*$ exchange which would diverge (properly taking into 
account the exchange of higher spin particles was actually one 
of the main motivations of our Regge model). Let's also notice that a recent
direct experimental determination of the proton electric form factor
at Jefferson Lab \cite{gep} clearly shows that the cut-off mass
needed to fit the large $Q^2$ data is not consistent with the cut-off mass
which is determined by the proton charge radius.

\begin{figure}[h]
\epsfxsize=7. cm
\epsfysize=9. cm
\centerline{\epsffile{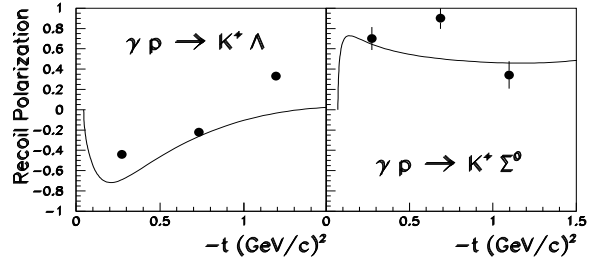}}
\vskip -3.5 cm
\caption[]{$t$ dependence of the $\Lambda$ and $\Sigma$ recoil
polarizations recently measured at Bonn in 
\underline{photoproduction}~\cite{bonn}.}
\label{fig:recoil}
\end{figure}

We now turn to polarization observables. We first show on 
Fig.~\ref{fig:recoil} that the 
$\Lambda$ and $\Sigma$ single recoil polarizations recently measured at Bonn~\cite{bonn} in photoproduction are reasonnably predicted 
by our model at forward angles (At larger angles, other contribution besides
$t$-channel exchanges are expected to contribute~: $u$-channel exchanges, 
resonances contributions at low energies, etc...). It is the interference between the $K$ and $K^*$ trajectories -arising from the $e^{i\alpha_{K,K^*}(t)}$ signature
term in the $K$ and $K^*$ Regge propagators, see Ref.~\cite{article}- which produces the non-zero polarization in the present model. The negative (positive) $\Lambda$ ($\Sigma$) recoil polarization is directly related to the relative sign of the $g_{K^*N\Lambda}$ couplings with respect to the $K$ ($g_{K\Lambda N}<0$ 
and $g_{K\Sigma N}>0$ whereas $g_{K^*(\Lambda,\Sigma)N}$ are both negative).\par
At Jefferson Lab, with electron beams, double polarization observables
will soon be accessed for the first time~\cite{carman}. These will put further 
stringent constraints on the models and will allow to disentangle the contributions of the $K$ and the $K^*$ exchange.
Typical behaviors for kinematics accessible at Jefferson Lab ($E_e$ = 6 GeV, 
$\theta_{e'}$ = 13$^{\circ}$, $E_{\gamma^*}$ = 2.643 GeV) are shown in
Fig.~\ref{fig:polar} for $\Phi$=0$^o$ and 180$^o$ (The  definitions and the  notations are given  in
Appendix A of Ref.~\cite{lag94}, and the $z$-axis is chosen along the direction
of the virtual photon).  If  only one Regge trajectory is retained the  induced
polarization  $P^{\circ}_Y$  (unpolarized  electrons)  vanishes:   it is
different from zero when the  two trajectories interfere. It is worth noting
that $P^{\circ}_Y$, which is the ``extension" to {\it electroproduction} of 
the {\it photoproduction} recoil polarization of Fig.~\ref{fig:recoil},
is now positive (at $\Phi$=0$^o$). Indeed, in photoproduction, this
observable is sensitive only to $\sigma_T^{Y}$ whereas in electroproduction
the additionnal $\sigma_L^{Y}$, $\sigma_{TT}^{Y}$ and 
$\sigma_{TL}^{Y}$ cross sections enter. These latter contributions
come with an oppposite sign with respect to $\sigma_T^{Y}$ here. The sideways
($P'_X$) and longitudinal ($P'_Z$) transfered polarizations indicate
the relative amount of $K$ and $K^*$ exchanges. The strong $\Phi$ dependence
evidenced on Fig.~\ref{fig:polar} can also be used to disentangle the
various contributions. At large transfers, 
it was noted that, in Deep Inelastic Scattering,
the strange quark can be used to follow the spin transfer from the probe 
to the emitted $\Lambda$. It will be interesting to compare such approaches
to our Regge calculations. 

\begin{figure}[h]
\epsfxsize=8. cm
\epsfysize=12. cm
\centerline{\epsffile{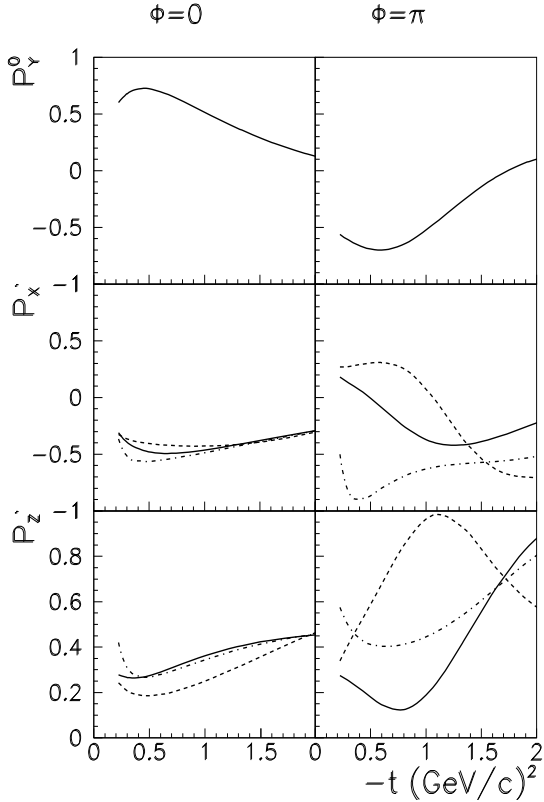}}
\caption[]{The $t$ dependence of the three $\Lambda$ polarizations which
do not vanish in coplanar kinematics, for $E_e$=6 GeV, $W$=2.2 GeV and $Q^2$=1
GeV$^2$. Dashed lines: $K$ exchange only. Dash-dotted lines: $K^*$
exchange only. Full lines: complete model. Left panels~: $\Phi$=0$^o$ degrees,
right panels~: $\Phi$=180$^o$}
\label{fig:polar}
\end{figure}

To conclude,  this simple  and elegant  Regge trajectory  exchange model
accounts fairly well for the whole set of available data and the  rather
accurate first measurement at Jefferson Lab.  Since the model depends on
very  few  parameters,  the  forthcoming  data from Jefferson Lab will
constitute  a  stringent  test (that  may  eventually call for a more
fundamental partonic description). Nevertheless, it provides already  a
good starting  point to  compute and  analyze strangeness  production in
nuclei.

We would like to acknowledge useful discussion with P. Bydzovsky. 
This work was supported in part by the French CNRS/IN2P3, the French
Commissariat \`a l'\'energie Atomique and the Deutsche 
Forschungsgemeinschaft (SFB443).


\begin{thebibliography}{99}
\bibitem{bobw}
R.A. Williams, C.R. Ji and S.R. Cotanch,
Phys. Rev. C {\bf 46}, 1617 (1992).
\bibitem{david}
J.C. David, C. Fayard, G.H. Lamot and B. Saghai,
Phys. Rev. C {\bf 53}, 2613 (1996).
\bibitem{paper}
M. Guidal, J.-M. Laget and M. Vanderhaeghen,
Phys. Lett. B {\bf 400}, 6. (1997).
\bibitem{article}
M. Guidal, J.-M. Laget and M. Vanderhaeghen,
Nucl. Phys. A {\bf 627}, 645 (1997).
\bibitem{electro}
M. Vanderhaeghen, M. Guidal and J.-M. Laget,
Phys. Rev. C {\bf 57}, 1454 (1998).
\bibitem{baker}
G. Niculescu et al.,
Phys. Rev. Lett. {\bf 81}, 1805 (1998).
\bibitem{bebek77b}
C.J. Bebek, C.N. Brown, R.V. Kline, F.M. Pipkin, S.W. Raither,
L.K. Sisterson, A. Browman, K.M. Hanson, D. Larson and A. Silverman,
Phys. Rev. D {\bf 15}, 3082 (1977).
\bibitem{brauel79}
P. Brauel, T. Canzler, D. Cords, R. Felst, G. Grindhammer, M. Helm,
W.-D. Kollmann, H. Krehbiel and M. Sch\"{a}dlich, 
Z. Phys. C {\bf 3}, 101 (1979).
\bibitem{bonn}
M.Q. Tran et al., Phys. Lett. B {\bf 445}, 20 (1998).
\bibitem{feller}
P. Feller, D. Menze, U. Opara, W. Schulz and W.J. Schwille,
Nucl. Phys. B {\bf 39}, 413 (1972).
\bibitem{bebek77}
C.J. Bebek, C.N. Brown, P. Bucksbaum, M. Herzlinger, S.D. Holmes,
C.A. Lichtenstein, F.M. Pipkin, S.W. Raither and L.K. Sisterson, 
Phys. Rev. D {\bf 15}, 594 (1977).
\bibitem{azemoon}
T. Azemoon, I. Dammann, C. Driver, D. L\"{u}ke, G. Specht,
K. Heinloth, H. Ackermann, E. Ganssauge, F. Janata and D. Schmidt,
Nucl. Phys. B {\bf 95}, 77 (1975).
\bibitem{brown}
C. N. Brown, C.R. Canizares, W.E. Cooper, A.M. Eisner, G.J. Feldman,
C.A. Lichtenstein, L.Litt, W. Lockeretz, V.B. Montana and F.M. Pipkin,
Phys. Rev. Lett. {\bf 28}, 1086 (1972).
\bibitem{bebek74}
C. J. Bebek et al.,
Phys. Rev. Lett. {\bf 32}, 21 (1974).
\bibitem{amen}
S. R. Amendolia et al.,
Phys. Lett. B {\bf 178}, 435 (1986).
\bibitem{amen2}
S. R. Amendolia et al.,
Phys. Lett. B {\bf 146}, 116 (1984).
\bibitem{gep}
M. K. Jones et al., nucl-ex/9910005 (1999).
\bibitem{carman}
JLab proposals, E99-006, D. Carman et al., E98-101, O. Baker et al. .
\bibitem{lag94} J.-M. Laget,
Nucl. Phys. A {\bf 579}, 333 (1994).
\end{thebibliography}
\end{document}